\documentclass[a4paper]{article}
\usepackage{ISCSLP2026}
\usepackage{ifthen}
\usepackage{comment}
\usepackage{amsmath,amssymb,amsfonts}
\usepackage{multirow}
\usepackage{array}
\usepackage{url}
\usepackage{algorithm}
\usepackage{algpseudocode}

\newboolean{blind}
\setboolean{blind}{false}

\title{AlphaFlowTSE: One-Step Generative Target Speaker Extraction via Conditional AlphaFlow}
\name{%
  \ifthenelse{\boolean{blind}}%
  {Anonymous to ISCSLP}%
  {Duojia Li$^{1,3}$, Shuhan Zhang$^{3,4}$, Zihan Qian$^2$, Wenxuan Wu$^5$,\\
   Shuai Wang$^{2,3,**}$, Qingyang Hong$^1$, Lin Li$^{1,**}$, Haizhou Li$^{3,4,**}$\thanks{$^{**}$ Corresponding authors.}}%
}
\address{%
  \ifthenelse{\boolean{blind}}%
  {Anonymous to ISCSLP}%
  {$^1$ Xiamen University, Xiamen, China\\
   $^2$ Nanjing University, Suzhou, China\\
   $^3$ Shenzhen Loop Area Institute, Shenzhen, China\\
   $^4$ The Chinese University of Hong Kong, Shenzhen, China\\
   $^5$ The Chinese University of Hong Kong, Shatin, N.T., Hong Kong SAR, China}%
}
\email{%
  \ifthenelse{\boolean{blind}}%
  {Anonymous to ISCSLP}%
  {liduojia@stu.xmu.edu.cn, shuaiwang@nju.edu.cn}%
}

\newcommand{\deltapct}[1]{{\scriptsize #1}}
\newcommand{\deltadb}[1]{{\scriptsize #1}}

\begin{document}

\maketitle

\begin{abstract}
Target speaker extraction (TSE) aims to recover the speech of a specified speaker from a multi-talker mixture using a short enrollment utterance from that speaker. Real conversational applications require both high extraction quality and low-latency inference across complex mixture conditions. Existing generative TSE approaches improve speech fidelity, but diffusion and flow-based models usually rely on multi-step sampling. One-step variants reduce iterative inference cost, yet often require estimating a trajectory coordinate related to the mixing ratio, introducing an additional dependency. We propose AlphaFlowTSE, which formulates TSE as finite-interval transport from the observed mixture to clean target speech. AlphaFlowTSE learns a conditional mean velocity along a mixture-to-target trajectory and uses a JVP-free AlphaFlow objective that combines local trajectory matching with interval-consistent teacher--student supervision. This parameterization supports predictor-free inference while remaining compatible with an optional MR-enabled configuration. At separator NFE$=1$, the MR-enabled configuration attains the highest PESQ, ESTOI, and SI-SDR among the compared one-step systems on Libri2Mix; under predictor-free inference on REAL-T PRIMARY, AlphaFlowTSE yields lower downstream ASR error rates and higher target-speaker similarity.
\end{abstract}

\section{Introduction}

In multi-talker scenarios such as online meetings, hands-free communication, and far-field conversations, users are usually interested in one participant, while competing speakers and ambient sounds form the interference. For such applications, a practical personalized speech front-end should extract the target speech reliably under realistic acoustic conditions while keeping inference efficient enough for interactive use~\cite{zmolikova23_tse_overview,alcalapadilla25_interspeech}. Target speaker extraction (TSE) addresses this need by using auxiliary speaker information to recover the desired speech from a mixture. This paper focuses on single-channel audio-enrollment TSE, where a short non-overlapped target-speaker utterance provides the speaker cue~\cite{wang19h_interspeech,zmolikova19_speakerbeam,ochiai19_interspeech}.

Most TSE systems follow a discriminative formulation: given the mixture and a target-speaker representation derived from enrollment, a neural network directly estimates a mask or waveform~\cite{delcroix19_interspeech,ge20_interspeech}. Progress in time-domain separation~\cite{luo19_convtasnet} and Transformer-based architectures~\cite{subakan21_sepformer} has substantially improved this paradigm. These systems nevertheless optimize a direct point estimate of target speech. Under strong interference, domain mismatch, or complex overlap, their outputs may still exhibit distortion, artifacts, or over-suppression~\cite{zmolikova23_tse_overview}. This leaves room for complementary formulations that impose additional structure on speaker-conditioned target recovery.

As a complementary training formulation, conditional generative modeling views TSE as transport from an initial speech representation to target speech: under the enrollment condition, the model updates the current state along a trajectory toward the desired speaker signal. Diffusion-based TSE methods have shown promising fidelity and naturalness~\cite{kamo23_interspeech,zhang24_ddtse}, but reverse sampling usually needs many network evaluations unless accelerated by fast solvers or distillation~\cite{lu22_dpmsolver,salimans22_progressivedistill}. Flow matching offers a deterministic transport alternative by learning a conditional velocity field~\cite{lipman22_flowmatching,liu23_rectifiedflow}, and FlowTSE adapts this idea to enrollment-conditioned extraction~\cite{navon25_interspeech}. However, conventional diffusion sampling and flow integration remain iterative, so their computational cost is tied to the number of function evaluations (NFE).

Finite-interval modeling provides a more direct path toward one-step generative TSE~\cite{hsieh25_adflowtse,shimizu25_meanflowtse}. Rather than explicitly integrating the whole trajectory, these methods learn an average velocity over a finite span, so the main update can use a single network evaluation. MeanFlow aligns this training target with the finite update used at inference~\cite{geng25_meanflows}, and related one-step mean-flow modeling has also been explored for speech enhancement~\cite{li26_meanflowse}, while AlphaFlow improves long-interval training with a Jacobian--vector product (JVP)-free objective that combines local trajectory matching and teacher--student interval consistency~\cite{zhang25_alphaflow}. In TSE, however, reliable one-step generation also depends on how the transport trajectory is chosen. If the finite interval is indexed by a mixing-ratio-related coordinate, inference requires estimating that coordinate first, and performance on real conversational mixtures can then depend on how well the auxiliary predictor transfers across domains.

Based on this observation, we propose AlphaFlowTSE, a one-step conditional generative framework with single-evaluation inference. Our key insight is to formulate TSE as mixture-to-target finite-interval transport: the observed mixture is the start state and clean target speech is the endpoint. This removes the mandatory dependence on an estimated mixing-ratio coordinate while retaining an optional MR-enabled configuration. AlphaFlowTSE learns an enrollment-conditioned mean velocity in the complex STFT domain and trains it with a JVP-free AlphaFlow objective, where local trajectory matching and interval-consistent teacher--student supervision align optimization with the finite update used at inference. The predictor-free path therefore extracts target speech directly from the observed mixture in one update.

We make three contributions. First, we recast TSE as mixture-to-target finite-interval transport, removing the mandatory dependence on trajectory-coordinate prediction while remaining compatible with optional MR-enabled operation. Second, we adapt JVP-free AlphaFlow training to enrollment-conditioned TSE, constraining conditional mean velocity with local trajectory matching and long-interval consistency. Third, experiments show that the MR-enabled configuration attains the highest PESQ, ESTOI, and SI-SDR among the compared one-step systems at NFE$=1$ on Libri2Mix; under predictor-free inference on REAL-T PRIMARY, AlphaFlowTSE obtains lower ASR error rates and higher target-speaker similarity.

\section{Background}
\label{sec:tse}

Generative TSE is moving from iterative conditional generation toward transport with few network evaluations. We review flow-based TSE as the basis for conditional velocity fields and trajectory transport, finite-interval mean velocity and the MR-indexed trajectories used by recent low-NFE baselines, and AlphaFlow's JVP-free interval-consistency principle.

\subsection{Flow-based Generative TSE}
\label{sec:flow_tse}

In single-channel TSE, we observe a mixture waveform $y\in\mathbb{R}^{L}$ and a short enrollment utterance $e\in\mathbb{R}^{L_e}$ that specifies the target speaker.
A common additive model is
\begin{equation}
y = s + b,
\label{eq:tse_mix}
\end{equation}
where $s$ is the clean target speech and $b$ aggregates all non-target components, including interfering speakers and background noise.
Given $(y,e)$, TSE estimates $\hat{s}$ while preserving target-speaker identity and suppressing $b$.
The enrollment $e$ is typically taken from non-overlapped speech to avoid contaminating the speaker cue~\cite{ochiai19_interspeech,zmolikova23_tse_overview}.

Conditional generative TSE casts extraction as transport in a representation space.
Let $z$ denote a speech representation, such as STFT features, and $c$ the mixture/enrollment condition.
Along a trajectory, $z_t$ moves from a start point to an endpoint while a neural network learns the conditional update rule.
When applied iteratively, this rule incurs a cost governed by the number of function evaluations (NFE), linking trajectory design to low-latency TSE.

Flow matching describes this transport by learning an instantaneous velocity field~\cite{lipman22_flowmatching}.
A standard construction linearly interpolates between a source sample $z_0\sim p_0$ and a target sample $z_1$:
\begin{equation}
z_t \triangleq (1-t)\,z_0 + t\,z_1,\qquad t\in[0,1].
\label{eq:fm_path}
\end{equation}
The model learns a velocity field $v_\theta(z_t,t,c)$ that describes state evolution through the ODE
\begin{equation}
\frac{d z_t}{d t} = v_\theta(z_t,t,c),
\label{eq:fm_ode}
\end{equation}
For the linear path in~\eqref{eq:fm_path}, the ground-truth velocity is the constant vector $(z_1-z_0)$, yielding the flow-matching objective
\begin{equation}
\mathcal{L}_{\mathrm{FM}}(\theta)
=
\mathbb{E}_{t,z_0,z_1}\Bigl[
\lVert v_\theta(z_t,t,c)-(z_1-z_0)\rVert_2^2
\Bigr].
\label{eq:fm_loss}
\end{equation}
Sampling integrates Eq.~\eqref{eq:fm_ode} from $t=0$ to $t=1$, so inference cost increases with the integration steps.
FlowTSE applies this paradigm to enrollment-conditioned TSE by conditioning the velocity field on mixture and enrollment cues and generating target speech along the learned trajectory~\cite{navon25_interspeech}.
This gives generative TSE a clear transport formulation, but standard flow transport remains iterative; reducing NFE close to $1$ requires direct long-interval modeling.

\subsection{One-Step Generative TSE}
\label{sec:onestep_meanvel}

One-step generation replaces many small integration steps with a single finite-interval update.
For low-latency TSE, the model learns the average velocity from the current state to the target endpoint, allowing the main extraction process to use one network evaluation.

This idea can be formalized by a mean-velocity model~\cite{geng25_meanflows}.
Let $0\le t<r\le 1$ denote two time endpoints on the same trajectory, with state $z_t$ at time $t$.
Under condition $c$, the mean-velocity network $u_\theta(z_t,t,r,c)$ predicts the finite-interval average velocity from $t$ to $r$, estimating the endpoint as
\begin{equation}
\hat z_r = z_t + (r-t)\,u_\theta(z_t,t,r,c).
\label{eq:bg_meanvel_jump}
\end{equation}
Setting $(t,r)=(0,1)$ directly gives one-step generation, i.e., NFE$=1$.

MeanFlow-TSE applies this finite-interval idea to conditional generative TSE, predicting the remaining transport from the current mixture-related state to the target endpoint in one update rather than by iterative integration~\cite{shimizu25_meanflowtse}.

A common trajectory choice in recent low-NFE TSE baselines is an MR-indexed background-to-target path~\cite{hsieh25_adflowtse,shimizu25_meanflowtse}.
Here, the background contains all non-target mixture components; let $Y$, $S$, and $B$ denote the STFT-domain mixture, target, and background, respectively.
In synthetic mixtures, the mixture can be approximately associated with a mixing-ratio-like coordinate $\tau^\star\in[0,1]$:
\begin{equation}
Y \approx (1-\tau^\star)B + \tau^\star S.
\label{eq:bg2tgt_mr_def}
\end{equation}
This gives a linear background-to-target path
\begin{equation}
x_\tau \triangleq (1-\tau)\,B + \tau\,S,\qquad \tau\in[0,1],
\label{eq:bg2tgt_path}
\end{equation}
with endpoints $x_0=B$ and $x_1=S$.
If $\tau^\star$ were known at inference, the system could start near the mixture location and traverse only the remaining span to $\tau=1$.
Because $\tau^\star$ is unavailable at test time, AD-FlowTSE estimates $\hat{\tau}$ from the mixture and enrollment and then transports from $\hat{\tau}$ to $1$ with a small number of updates~\cite{hsieh25_adflowtse}; MeanFlow-TSE combines this MR-indexed trajectory with the finite-interval update in~\eqref{eq:bg_meanvel_jump} for one-step extraction~\cite{shimizu25_meanflowtse}.
Inference therefore depends on estimating an additional trajectory coordinate, introducing a prediction module whose cross-domain transfer can affect real conversational mixtures.

Beyond trajectory choice, one-step generation also faces a training-stability challenge.
When inference is compressed into a single long-interval update, the model must predict accurate average velocities over long spans while remaining consistent across different $(t,r)$ choices.
Directly enforcing this consistency often involves time derivatives of the model output or JVPs, increasing overhead and potentially destabilizing optimization when supervision signals interact~\cite{geng25_meanflows}.
AlphaFlow addresses this requirement through JVP-free interval-consistency training while preserving the one-step update in~\eqref{eq:bg_meanvel_jump}.

\begin{figure*}[t]
  \centering
  \includegraphics[width=0.98\textwidth]{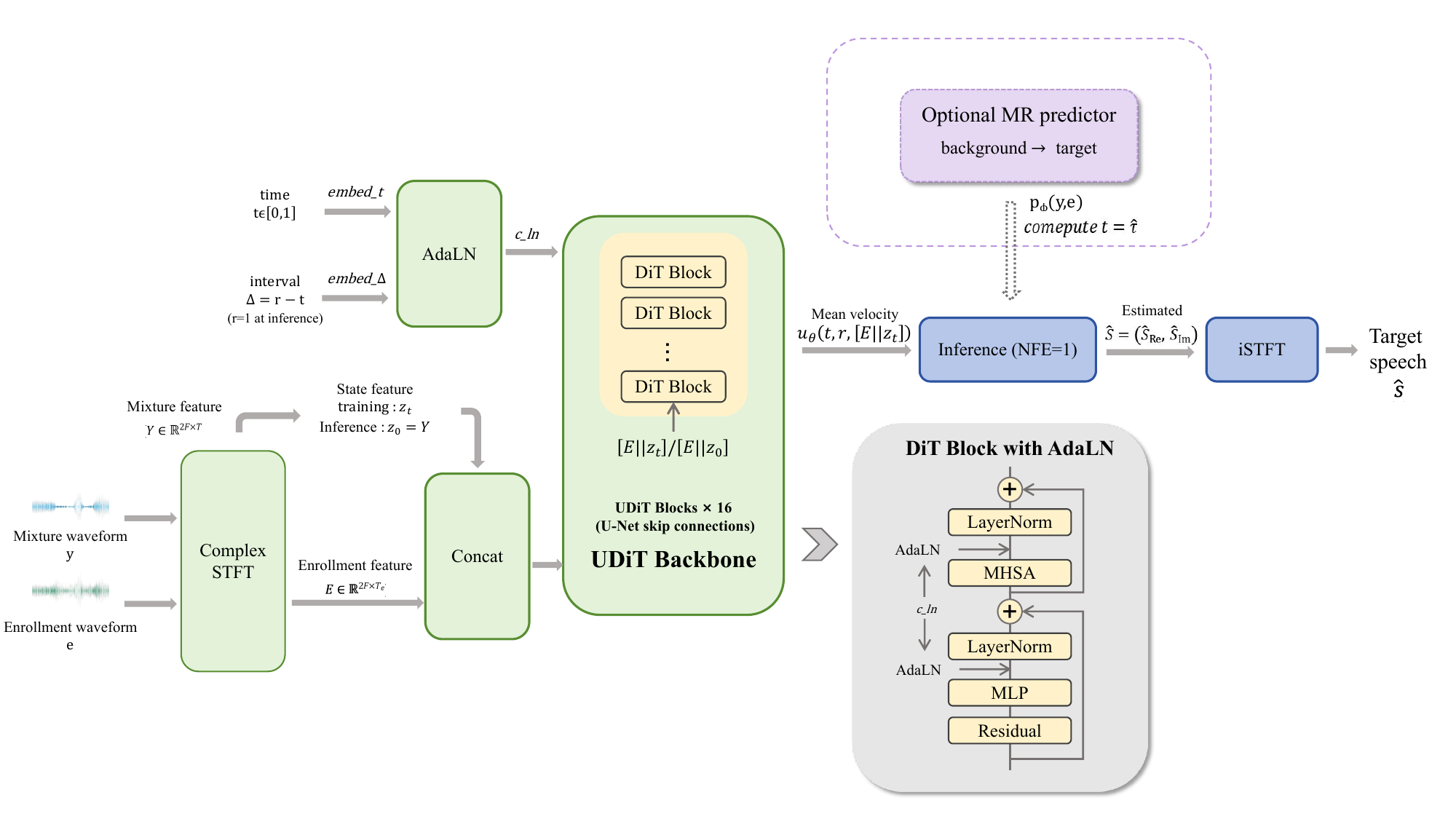}
  \caption{Overall architecture of AlphaFlowTSE.
  Given a mixture waveform $y$ and an enrollment utterance $e$, we compute complex STFT features and form the mixture feature $Y$ and enrollment feature $E$ (real/imaginary concatenation).
  During training, the backbone takes the current state feature $z_t$; during inference we initialize $z_0=Y$. The enrollment feature is concatenated as a temporal prefix, yielding $[E\!\parallel\! z_t]$ (or $[E\!\parallel\! z_0]$ at inference), which is fed to the UDiT backbone.
  The backbone is conditioned via AdaLN on the absolute time $t$ and the interval length $\Delta=r-t$ (with $r=1$ at inference), and predicts the mean velocity for finite-interval transport, denoted $u_\theta(t,r,[E\!\parallel\! z_t])$.
  One-step inference (\texttt{NFE}$=1$) produces an estimated complex STFT $\hat{S}=(\hat{S}_{\mathrm{Re}},\hat{S}_{\mathrm{Im}})$, which is converted to the target waveform $\hat{s}$ by iSTFT.
  The dashed module is an optional mixing-ratio predictor used by the MR-enabled background-to-target configuration to predict the start coordinate $\hat{\tau}$.}
  \label{fig:alphaflowtse_arch}
\end{figure*}

\subsection{AlphaFlow}
\label{sec:alphaflow}

AlphaFlow trains finite-interval mean-velocity models without explicit JVP computation~\cite{zhang25_alphaflow}.
It combines a trajectory-matching signal that anchors predictions to the intended transport direction with an interval-consistency signal that encourages consistent average-velocity predictions across interval lengths and starting points.
Unlike methods that differentiate model outputs with respect to time, AlphaFlow builds the consistency target through a stop-gradient teacher--student construction.

Given an interval $(t,r)$, AlphaFlow introduces an intermediate time point
\begin{equation}
s = \alpha r + (1-\alpha)t,\qquad 0<\alpha\le 1,
\label{eq:alphaflow_s}
\end{equation}
where $\alpha$ controls the position of the intermediate point between the start $t$ and endpoint $r$.
A teacher prediction is evaluated at the intermediate state $z_s$ over $(s,r)$ with stop-gradient.
The teacher prediction and the direct trajectory anchor jointly supervise the student prediction $u_\theta(z_t,t,r,c)$, so the longer-interval student is constrained by a shorter sub-interval teacher.

The parameter $\alpha$ also controls training difficulty.
Values close to $1$ favor direct trajectory matching, while smaller values strengthen teacher-guided consistency over longer spans.
AlphaFlow therefore anneals $\alpha$ from values near $1$ to smaller values, gradually shifting toward stronger interval-consistency constraints~\cite{zhang25_alphaflow}.

This construction fits the mixture-to-target trajectory adopted later in this paper.
Because both $Y$ and $S$ are available during training, intermediate states can be obtained by closed-form interpolation rather than model-generated samples.
The teacher branch therefore remains on the true training trajectory and provides on-trajectory supervision for one-step mean-velocity learning.

\section{Method}

As illustrated in Figure~\ref{fig:alphaflowtse_arch}, AlphaFlowTSE formulates target speaker extraction as a conditional finite-interval transport problem in the spectral domain.
Given a mixture and an enrollment utterance, we first define a deterministic mixture-to-target trajectory from the observed mixture to clean target speech, then parameterize finite-interval transport with an enrollment-conditioned mean-velocity network, and train it with a JVP-free AlphaFlow objective.
We instantiate AlphaFlow interval consistency on a deterministic mixture-to-target trajectory fully specified by each training pair $(Y,S)$.
The trajectory provides closed-form intermediate states on the true training path, and the resulting teacher--student target requires neither explicit JVP computation, model rollout, nor a separate intermediate-state estimator.
At inference, the model starts from the observed mixture, performs one finite-interval update, and reconstructs the target waveform by iSTFT (NFE$=1$).

\subsection{AlphaFlowTSE}
\label{sec:alphaflowtse}

AlphaFlowTSE learns an enrollment-conditioned one-step transport that maps an observed mixture to target speech.
We operate in the complex STFT domain and represent each spectrum by concatenating real and imaginary parts along the channel axis.
Omitting the batch dimension, we denote the mixture and target spectra as $Y,S\in\mathbb{R}^{2F\times T}$, where $F$ and $T$ are the numbers of frequency bins and time frames, and the enrollment spectrum as $E\in\mathbb{R}^{2F\times T_e}$.

\noindent\textbf{Mixture-to-target trajectory.}
We define a linear trajectory from the mixture to target speech in the STFT domain:
\begin{equation}
z_t \triangleq (1-t)\,Y + t\,S,\qquad t\in[0,1].
\label{eq:aftse_path}
\end{equation}
The trajectory starts from the observed mixture, $z_0=Y$, and ends at clean target speech, $z_1=S$.
For any forward interval $0\le t\le r\le 1$, the mean-velocity network $u_\theta(z_t,t,r;E)$ predicts the average velocity from $t$ to $r$, and the endpoint is estimated as
\begin{equation}
\hat z_r = z_t + (r-t)\,u_\theta(z_t,t,r;E).
\label{eq:aftse_jump}
\end{equation}
At inference, we fix $(t,r)=(0,1)$ and apply one update from the observed mixture:
\begin{equation}
\hat S = Y + u_\theta(Y,0,1;E),
\label{eq:aftse_onestep}
\end{equation}
Equation~\eqref{eq:aftse_onestep} performs a residual-form finite-interval update from the observed mixture to the target endpoint. During training, $u_\theta$ is evaluated at intermediate trajectory states and variable intervals and optimized through local velocity matching and interval-consistency supervision.
This training exposes the model to both local transport directions and long-span updates, while Eq.~\eqref{eq:aftse_onestep} remains the endpoint update used at inference.
The predicted spectrum is then converted to a waveform by iSTFT reconstruction.
The trajectory in~\eqref{eq:aftse_path} is used only to construct training-time supervision from paired $(Y,S)$; at test time, $S$ is unknown and the update in~\eqref{eq:aftse_onestep} depends only on $(Y,E)$.

\noindent\textbf{Network parameterization.}
We implement $u_\theta$ with a U-Net style Diffusion Transformer backbone (UDiT)~\cite{shimizu25_meanflowtse}.
The network input is formed by concatenating the enrollment spectrum $E$ as a temporal prefix to the current state $z_t$; the output tokens corresponding to the mixture segment are interpreted as the predicted mean velocity.
To support interval-dependent prediction with a single model, we condition each DiT block on the start time $t$ and the interval length $\Delta=r-t$ via adaptive layer normalization, using a conditioning vector $c_{t,\Delta}=\mathrm{emb}(t)+\mathrm{emb}(\Delta)$.

\subsection{JVP-Free AlphaFlow Training}
\label{sec:alphaflowtse_objective}
\noindent\textbf{Training objective.}
We train $u_\theta$ for accurate long-interval prediction and consistency across interval choices.
For the linear interpolation path in~\eqref{eq:aftse_path}, the trajectory velocity is constant:
\begin{equation}
v \triangleq \frac{d z_t}{dt} = S - Y.
\label{eq:aftse_vtrue}
\end{equation}
Thus, $v$ provides a direct trajectory-matching anchor.
One-step inference spans the long interval from $t=0$ to $r=1$, while local trajectory matching does not explicitly enforce consistency across different $(t,r)$ queries.
We therefore add the teacher--student interval-consistency objective of JVP-free AlphaFlow.

The deterministic mixture-to-target trajectory allows any intermediate state $z_s$ to be computed exactly from $(Y,S)$ by linear interpolation.
The teacher branch therefore does not require model rollout or an auxiliary intermediate-state predictor; it is a stop-gradient prediction on a shorter remaining interval $(s,r)$ along the same true trajectory.
The student starts from the earlier state $z_t$ and learns the average velocity over the longer interval $(t,r)$.
Both predictions are therefore evaluated at analytically defined states from the same paired trajectory.
Teacher--student consistency directly constrains long- and short-interval predictions along the same mixture-to-target trajectory.

For a residual tensor $D\in\mathbb{R}^{2F\times T}$, we define the per-sample mean squared error as
\begin{equation}
m(D) \triangleq \frac{1}{2FT}\,\lVert D \rVert_F^2,
\label{eq:aftse_mse}
\end{equation}
where $\lVert\cdot\rVert_F$ denotes the Frobenius norm.

\noindent\textbf{Local matching anchor.}
We train the velocity predictor on the diagonal slice $r=t$ by matching its output to the true trajectory velocity.
At $r=t$, the interval displacement is zero, while $u_\theta(z_t,t,r;E)$ remains defined as a velocity; regressing to $v$ at $\Delta=0$ provides local direction supervision.
We sample $t\in(0,1)$, set $r=t$, and compute $z_t$ from~\eqref{eq:aftse_path}.
Let $D_{\mathrm{FM}} \triangleq u_\theta(z_t,t,t;E)-v$.
We denote $\mathrm{sg}(\cdot)$ as the stop-gradient operator, and apply an adaptive weighting
\begin{equation}
\ell_{\mathrm{adp}}(D)
\triangleq
\mathrm{sg}\Bigl(\bigl(m(D)+\epsilon_{\mathrm{adp}}\bigr)^{\gamma-1}\Bigr)\,m(D).
\end{equation}
where $\epsilon_{\mathrm{adp}}>0$ is a small constant, $\gamma\in[0,1]$, and $\mathrm{sg}(\cdot)$ stops gradients through its argument.
The corresponding objective is
\begin{equation}
\mathcal{L}_{\mathrm{FM}}(\theta)
=
\mathbb{E}_{t}\Bigl[\ell_{\mathrm{adp}}(D_{\mathrm{FM}})\Bigr].
\label{eq:aftse_lfm}
\end{equation}

\noindent\textbf{AlphaFlow interval consistency.}
To constrain long-interval prediction without computing JVPs, we sample $0\le t<r\le 1$, choose $\alpha\in(0,1]$, and define an intermediate time and state:
\begin{equation}
\begin{aligned}
s &\triangleq \alpha r + (1-\alpha)t,\\
z_s &\triangleq (1-s)\,Y + s\,S.
\end{aligned}
\label{eq:aftse_s}
\end{equation}
Here $s$ lies inside $(t,r)$, and $\alpha$ controls the length of the teacher sub-interval.
When $\alpha$ is close to $1$, $s$ approaches $r$ and the target is closer to direct trajectory matching; as $\alpha$ decreases, $s$ moves toward $t$ and the teacher branch imposes a stronger long-interval consistency constraint.

Given $(t,r,s)$, the student prediction is $u_\theta(z_t,t,r;E)$ and the teacher prediction is
\[
\tilde{u} \triangleq \mathrm{sg}\bigl(u_\theta(z_s,s,r;E)\bigr).
\]
The student and teacher use the same mean-velocity network, but gradients through the teacher branch are blocked by $\mathrm{sg}(\cdot)$; thus this is not an additional EMA teacher.
Because $z_s$ is computed in closed form from $(Y,S)$, the teacher prediction is evaluated on the sub-interval $(s,r)$ of the true mixture-to-target trajectory.
The AlphaFlow target velocity is then
\begin{equation}
u^\star_{\alpha}
\triangleq
\alpha\,v + (1-\alpha)\,\tilde u ,
\label{eq:aftse_target}
\end{equation}
where $\alpha v$ keeps the target anchored to the true transport direction, while $(1-\alpha)\tilde u$ introduces consistency from the shorter remaining interval.
We denote the residual by $D_{\mathrm{MF}} \triangleq u_\theta(z_t,t,r;E)-u^\star_{\alpha}$.
To balance sample contributions under different $\alpha$, we adopt a bounded reweighting:
\begin{equation}
\ell_{\mathrm{bnd}}(D;\alpha)
\triangleq
\mathrm{sg}\left(\frac{\kappa}{m(D)+\alpha\kappa+\varepsilon}\right)\,m(D).
\label{eq:aftse_bnd}
\end{equation}
where $\kappa>0$ controls saturation and $\varepsilon>0$ is a numerical constant.
The corresponding interval objective is
\begin{equation}
\mathcal{L}_{\mathrm{MF}}(\theta)
=
\mathbb{E}_{t,r}\Bigl[\ell_{\mathrm{bnd}}(D_{\mathrm{MF}};\alpha)\Bigr].
\label{eq:aftse_lmf}
\end{equation}

We use decoupled training: each sample is assigned to the FM anchor branch with probability $\rho$ and to the MF consistency branch otherwise.
Equivalently, the expected objective is
\begin{equation}
\mathcal{L}(\theta)
=
\rho\,\lambda_{\mathrm{FM}}\,\mathcal{L}_{\mathrm{FM}}(\theta)
+
(1-\rho)\,\lambda_{\mathrm{MF}}\,\mathcal{L}_{\mathrm{MF}}(\theta),
\label{eq:aftse_total}
\end{equation}
with constants $\lambda_{\mathrm{FM}},\lambda_{\mathrm{MF}}>0$.
During training, $\alpha$ is annealed from $1$ to a floor value $\alpha_{\min}$, gradually shifting the supervision from pure trajectory matching toward teacher-guided interval consistency~\cite{zhang25_alphaflow}.

Algorithm~\ref{alg:alphaflowtse_training} summarizes JVP-free AlphaFlowTSE training on the mixture-to-target path.
In the FM branch, the model learns the true trajectory velocity $v=S-Y$.
In the MF branch, the student predicts the long-interval average velocity from $z_t$, while the teacher predicts a shorter remaining-interval velocity from the closed-form state $z_s$ and enters the target with stop-gradient.
Thus, training requires neither explicit JVPs nor intermediate states generated by the model or an extra module.

\begin{algorithm}[t]
\caption{AlphaFlowTSE: Training (mixture-to-target).}
\label{alg:alphaflowtse_training}
\footnotesize
\begin{algorithmic}[1]
\Require Training batch $(Y,S,E)$; branch probability $\rho$; weights $\lambda_{\mathrm{FM}},\lambda_{\mathrm{MF}}$; $\alpha$-schedule params $(k_s,k_e,\gamma_\alpha,\alpha_{\min})$; bounded-loss params $(\kappa,\varepsilon)$.
\For{training iteration $k=1,2,\dots$}
    \State $\alpha \gets \mathrm{AlphaSchedule}(k; k_s,k_e,\gamma_\alpha,\alpha_{\min})$ \Comment{from $1$ to $\alpha_{\min}$}
    \State Sample $q\sim\mathrm{Bernoulli}(\rho)$ \Comment{switch between FM and MF}
    \State $v \gets S - Y$ \Comment{true path velocity}
    \If{$q=1$} \Comment{Local trajectory matching (FM)}
        \State Sample $t\in(0,1)$ and set $r\gets t$
        \State $z_t \gets (1-t)Y + tS$
        \State $u \gets u_\theta(z_t,t,r;E)$
        \State $\mathcal{L} \gets \lambda_{\mathrm{FM}}\,\ell_{\mathrm{adp}}(u-v)$
    \Else \Comment{Interval consistency (JVP-free AlphaFlow)}
        \State Sample $(t,r)$ with $0\le t<r\le 1$ (including long spans)
        \State $s \gets \alpha r + (1-\alpha)t$
        \State $z_t \gets (1-t)Y + tS$, \ \ $z_s \gets (1-s)Y + sS$
        \State $u \gets u_\theta(z_t,t,r;E)$ \Comment{student}
        \State $\tilde u \gets \mathrm{sg}\bigl(u_\theta(z_s,s,r;E)\bigr)$ \Comment{teacher (stop-gradient)}
        \State $u^\star_\alpha \gets \alpha v + (1-\alpha)\tilde u$
        \State $\mathcal{L} \gets \lambda_{\mathrm{MF}}\,\ell_{\mathrm{bnd}}(u-u^\star_\alpha;\alpha)$
    \EndIf
    \State Update $\theta$ by gradient descent on $\mathcal{L}$
\EndFor
\end{algorithmic}
\end{algorithm}

\subsection{Inference}
\label{sec:inference}

At test time, we compute complex STFTs using the same analysis parameters as in training and concatenate real and imaginary parts.
For the proposed mixture-to-target parameterization, the model starts from the observed mixture ($t=0$), applies the one-step update in~\eqref{eq:aftse_onestep} to the target endpoint ($r=1$), and reconstructs the waveform by iSTFT.
No iterative sampling, guidance, or refinement is used (NFE$=1$).

In practice, training is performed on fixed-duration segments whereas test utterances may be longer.
For long utterances, we process the mixture spectrogram in contiguous time chunks, apply the one-step estimator with the same enrollment condition to each chunk, and concatenate the predicted spectrograms along the time axis before waveform reconstruction.

\noindent\textbf{Optional MR-enabled configuration.}
AlphaFlowTSE also supports a background-to-target configuration for controlled comparison with MR-indexed baselines.
Let $B\in\mathbb{R}^{2F\times T}$ denote the background/interference spectrum available during training and define
\begin{equation}
x_\tau \triangleq (1-\tau)\,B + \tau\,S,\qquad \tau\in[0,1].
\label{eq:aftse_bt_path}
\end{equation}
This variant uses the same training objective, replacing the trajectory velocity with $v_{\mathrm{bg}} \triangleq S-B$ and the teacher state with $x_s=(1-s)B+sS$.
At test time, a separate regressor $p_\phi$ estimates the coordinate of the mixture on the background-to-target path:
\begin{equation}
\hat\tau = \sigma\bigl(p_\phi(y,e)\bigr), \qquad \hat\tau\in(0,1).
\label{eq:infer_tau}
\end{equation}
We then perform one jump from the observed mixture state with $t=\hat\tau$ and $r=1$:
\begin{equation}
\hat S = Y + (1-\hat\tau)\,u_\theta\bigl(Y,\hat\tau,1;E\bigr).
\label{eq:infer_mr_onestep}
\end{equation}
This configuration requires one additional forward pass for $p_\phi$. The predictor-free mixture-to-target path omits this step, and both configurations otherwise use the same reconstruction pipeline.

\section{Experiment}

\subsection{Datasets and Data Preparation}
\label{sec:datasets}

We train and benchmark the models on Libri2Mix from the LibriMix corpus~\cite{cosentino20_librimix} for controlled comparison with recent generative TSE systems, then evaluate zero-shot transfer on the REAL-T PRIMARY subset~\cite{li25da_interspeech}.
Unless otherwise specified, we use the same Libri2Mix configuration and SpeakerBeam-style informed data protocol as AD-FlowTSE and MeanFlow-TSE~\cite{hsieh25_adflowtse,shimizu25_meanflowtse,ochiai19_interspeech}.

\noindent\textbf{Libri2Mix dataset.}
We follow the official LibriMix recipe and use Libri2Mix (\texttt{min}, 16~kHz) under \texttt{clean} and \texttt{noisy} conditions, with WHAM! noise added in the latter following the LibriMix procedure~\cite{cosentino20_librimix,panayotov15_librispeech,wichern19_wham}.

In the informed TSE setting, each mixture is paired with a designated target source, and a short target-only enrollment segment is provided as the speaker condition.
We use the same mixture/enrollment file lists and metadata organization as AD-FlowTSE and MeanFlow-TSE~\cite{ochiai19_interspeech,hsieh25_adflowtse,shimizu25_meanflowtse}.

\noindent\textbf{REAL-T (real conversational mixtures).}
To evaluate transfer beyond synthetic mixtures, we further test on REAL-T~\cite{li25da_interspeech}.
REAL-T is a conversation-centric TSE benchmark built from real multi-speaker recordings.
It contains natural conversational overlap and enrollment utterances from non-overlapping regions of the same speaker.
We use the official PRIMARY evaluation subset and data release from the REAL-T project repository~\cite{li25da_interspeech}.\footnote{\url{https://github.com/REAL-TSE/REAL-T}}

\subsection{Experimental Setup}

For comparability with recent one-step TSE baselines, we match the front-end features, data format, long-utterance handling, and test procedure of AD-FlowTSE and MeanFlow-TSE~\cite{hsieh25_adflowtse,shimizu25_meanflowtse}.

\noindent\textbf{Front-end and model.}
All systems use complex STFT features from 16~kHz audio with $N_{\mathrm{FFT}}=510$ and hop size $H=128$.
Concatenating real and imaginary parts along the channel dimension yields $2F=512$ input channels.
Our separator follows the MeanFlow-TSE UDiT backbone~\cite{shimizu25_meanflowtse}, with 16 Transformer blocks, 16 attention heads, and hidden width 1024.
The enrollment spectrogram is concatenated with the mixture features as a temporal prefix and processed by the same network.

\noindent\textbf{Training setting.}
We train separate models for Libri2Mix \texttt{clean} and \texttt{noisy}, using random 3~s crops of mixture/target and enrollment signals as in the baseline protocol.
Optimization uses AdamW with bfloat16 mixed precision and gradient clipping (max norm 0.5).
Models are trained for 150 epochs with an initial learning rate of $2\times10^{-5}$, a short linear warmup, and cosine decay.
Training is conducted with distributed data-parallel on 8 NVIDIA H100 GPUs; we use a per-GPU batch size of 42 with gradient accumulation of 2 steps.

AD-FlowTSE and MeanFlow-TSE report long training cycles, up to 2000 epochs, to obtain their strongest checkpoints.
To reduce computation while retaining the released baseline initialization, we load only the network parameters from the public AD-FlowTSE checkpoints~\cite{hsieh25_adflowtse} and discard optimizer and scheduler states.
We restart optimization with the schedule above under our mixture-to-target AlphaFlowTSE objective.
As a sanity check, continuing AD-FlowTSE from the same checkpoint under its original objective produced no consistent gains under the same compute budget.

Time pairs $(t,r)$ for the mean-flow branch are sampled using the logit-normal strategy in AlphaFlow~\cite{zhang25_alphaflow}, with $(\mu,\sigma)=(-0.4,1.0)$.
To better match one-step inference, we additionally draw 15\% of samples from a large-span subset with $t\le 0.15$ and $r\ge 0.85$.
We anneal $\alpha$ from 1 to $\alpha_{\min}=0.1$ with a sigmoid schedule (epochs 5--100, $k=15$).
The FM and MF branches are sampled with equal probability and use weights $\lambda_{\mathrm{FM}}=0.6$ and $\lambda_{\mathrm{MF}}=0.4$.

\noindent\textbf{Inference and MR-predictor comparison.}
We evaluate two operating configurations. The predictor-free mixture-to-target configuration transports directly from the observed mixture to the target endpoint with one separator evaluation (NFE$=1$), using the baseline reconstruction and long-utterance protocol without iterative refinement~\cite{hsieh25_adflowtse,shimizu25_meanflowtse}.

The optional MR-enabled configuration uses the background-to-target trajectory for controlled comparison with MR-indexed baselines. Following the mixing-ratio predictor design in AD-FlowTSE/MeanFlow-TSE, a separate regressor $p_\phi(\cdot)$ estimates the time coordinate $\hat\tau\in(0,1)$ of the observed mixture on this trajectory.
We implement $p_\phi$ with an ECAPA-TDNN encoder~\cite{desplanques20_interspeech} and an MLP regression head, and apply SpecAugment during training~\cite{park19_interspeech}.
At inference, $p_\phi$ is evaluated once to obtain $\hat\tau$, followed by one separator transport from $t=\hat\tau$ to $r=1$ using the same reconstruction pipeline.

\begin{table*}[t]
  \centering
  \caption{Libri2Mix benchmark results (\texttt{min}, 16~kHz). $\uparrow$ indicates higher is better.
  DNSMOS OVRL denotes the DNSMOS-P.835 overall score, and SpkSim denotes speaker similarity.
  Results of prior systems are taken from the literature~\cite{navon25_interspeech,kamo23_interspeech,zhang24_ddtse,ku25_srssl,wang25_solospeech,hsieh25_adflowtse,shimizu25_meanflowtse}.
  The AlphaFlowTSE row reports the MR-enabled configuration for protocol alignment with prior MR-indexed one-step systems; predictor-free results are reported in Table~\ref{tab:librimix_mr_effect}.}
  \label{tab:librimix_overall}
  \footnotesize
  \setlength{\tabcolsep}{2.4pt}
  \renewcommand{\arraystretch}{1.08}
  \begin{tabular}{lccccc@{\hskip 6pt}ccccc}
    \toprule
    Method &
    \multicolumn{5}{c}{Libri2Mix Clean} &
    \multicolumn{5}{c}{Libri2Mix Noisy} \\
    \cmidrule(lr){2-6}\cmidrule(lr){7-11}
    & PESQ$\uparrow$ & ESTOI$\uparrow$ & SI-SDR$\uparrow$ & OVRL$\uparrow$ & SpkSim$\uparrow$
    & PESQ$\uparrow$ & ESTOI$\uparrow$ & SI-SDR$\uparrow$ & OVRL$\uparrow$ & SpkSim$\uparrow$ \\
    \midrule
    Mixture
      & 1.15 & 0.54 & 0.00  & 2.65 & 0.54
      & 1.08 & 0.40 & -1.93 & 1.63 & 0.46 \\
    \midrule
    DiffSep+SV~\cite{zhang24_ddtse}
      & 1.85 & 0.79 & --    & 3.14 & 0.83
      & 1.32 & 0.60 & --    & 2.78 & 0.62 \\
    DDTSE~\cite{zhang24_ddtse}
      & 1.79 & 0.78 & --    & \textbf{3.30} & 0.73
      & 1.60 & 0.71 & --    & 3.28 & 0.71 \\
    DiffTSE~\cite{kamo23_interspeech}
      & 3.08 & 0.80 & 11.28 & --   & --
      & --   & --   & --    & --   & -- \\
    FlowTSE~\cite{navon25_interspeech}
      & 2.58 & 0.84 & --    & 3.27 & 0.90
      & 1.86 & 0.75 & --    & \textbf{3.30} & 0.83 \\
    SR-SSL~\cite{ku25_srssl}
      & 2.99 & --   & 16.00 & --   & --
      & --   & --   & --    & --   & -- \\
    SoloSpeech~\cite{wang25_solospeech}
      & --   & --   & --    & --   & --
      & 1.89 & 0.78 & 11.12 & --   & -- \\
    \midrule
    AD-FlowTSE~\cite{hsieh25_adflowtse}
      & 2.89 & 0.90 & 17.49 & 3.15 & \textbf{0.95}
      & 2.15 & 0.81 & 12.70 & 3.11 & \textbf{0.87} \\
    MeanFlowTSE~\cite{shimizu25_meanflowtse}
      & 3.26 & 0.93 & 18.80 & 3.21 & 0.92
      & 2.21 & 0.82 & 12.85 & 3.17 & 0.73 \\
    AlphaFlowTSE (ours)
      & \textbf{3.27} & \textbf{0.94} & \textbf{19.17} & 3.24 & 0.93
      & \textbf{2.28} & \textbf{0.85} & \textbf{13.16} & 3.19 & 0.76 \\
    \bottomrule
  \end{tabular}
\end{table*}

\begin{table*}[t]
  \centering
  \caption{Comparison of predictor-free and MR-enabled configurations on Libri2Mix (NFE$=1$). $\uparrow$ indicates higher is better.
  ``w/'' and ``w/o'' denote the MR-enabled and predictor-free configurations, respectively; signed relative-change columns compare \texttt{w/o} against \texttt{w/}. Best values are in bold.}
  \label{tab:librimix_mr_effect}
  \footnotesize
  \setlength{\tabcolsep}{1.2pt}
  \renewcommand{\arraystretch}{1.08}
  \begin{tabular}{
    >{\centering\arraybackslash}p{0.90cm}
    >{\centering\arraybackslash}p{2.25cm}
    c
    cccc
    cccc
    ccc
  }
    \toprule
    \multirow{2}{*}{Split} &
    \multirow{2}{*}{Method} &
    \multirow{2}{*}{MR pred.} &
    \multicolumn{4}{c}{DNSMOS} &
    \multirow{2}{*}{PESQ$\uparrow$} &
    \multirow{2}{*}{ESTOI$\uparrow$} &
    \multirow{2}{*}{SI-SDR$\uparrow$} &
    \multirow{2}{*}{SpkSim$\uparrow$} &
    \multicolumn{3}{c}{Relative change (\texttt{w/o} vs.\ \texttt{w/})} \\
    \cmidrule(lr){4-7}\cmidrule(lr){12-14}
    & & &
    SIG$\uparrow$ & BAK$\uparrow$ & OVRL$\uparrow$ & P808$\uparrow$ &
    & & & &
    $\Delta$OVRL(\%) & $\Delta$PESQ(\%) & $\Delta$SI-SDR (dB) \\
    \midrule

    \multirow[c]{6}{*}{\texttt{clean}}
      & \multirow[c]{2}{*}{\parbox[c]{2.25cm}{\centering AD-FlowTSE~\cite{hsieh25_adflowtse}}}
      & w/o
        & --   & --   & 3.02 & 3.44
        & 2.33 & 0.82 & 12.54 & 0.92
        & \multirow[c]{2}{*}{\deltapct{-4.1\%}}
        & \multirow[c]{2}{*}{\deltapct{-19.4\%}}
        & \multirow[c]{2}{*}{\deltadb{-4.95}} \\
      &  & w/
        & 3.47 & 3.90 & 3.15 & 3.59
        & 2.89 & 0.90 & 17.49 & \textbf{0.95} \\
      \addlinespace[0.2em]

      & \multirow[c]{2}{*}{\parbox[c]{2.25cm}{\centering MeanFlowTSE~\cite{shimizu25_meanflowtse}}}
      & w/o
        & 3.31 & 3.37 & 2.77 & 3.31
        & 1.53 & 0.53 & -6.00 & 0.60
        & \multirow[c]{2}{*}{\deltapct{-13.7\%}}
        & \multirow[c]{2}{*}{\deltapct{-53.1\%}}
        & \multirow[c]{2}{*}{\deltadb{-24.80}} \\
      &  & w/
        & 3.51 & 3.95 & 3.21 & 3.69
        & 3.26 & 0.93 & 18.80 & 0.92 \\
      \addlinespace[0.2em]

      & \multirow[c]{2}{*}{\parbox[c]{2.25cm}{\centering AlphaFlowTSE}}
      & w/o
        & 3.52 & 3.94 & 3.16 & 3.61
        & 3.04 & 0.92 & 18.50 & 0.92
        & \multirow[c]{2}{*}{\textbf{\deltapct{-2.5\%}}}
        & \multirow[c]{2}{*}{\textbf{\deltapct{-7.0\%}}}
        & \multirow[c]{2}{*}{\textbf{\deltadb{-0.67}}} \\
      &  & w/
        & \textbf{3.54} & \textbf{4.02} & \textbf{3.24} & \textbf{3.72}
        & \textbf{3.27} & \textbf{0.94} & \textbf{19.17} & 0.93 \\

    \midrule

    \multirow[c]{6}{*}{\texttt{noisy}}
      & \multirow[c]{2}{*}{\parbox[c]{2.25cm}{\centering AD-FlowTSE~\cite{hsieh25_adflowtse}}}
      & w/o
        & --   & --   & 2.87 & 3.23
        & 1.73 & 0.72 & 9.40  & 0.84
        & \multirow[c]{2}{*}{\deltapct{-7.7\%}}
        & \multirow[c]{2}{*}{\deltapct{-19.5\%}}
        & \multirow[c]{2}{*}{\deltadb{-3.30}} \\
      &  & w/
        & 3.43 & 3.92 & 3.11 & 3.48
        & 2.15 & 0.81 & 12.70 & \textbf{0.87} \\
      \addlinespace[0.2em]

      & \multirow[c]{2}{*}{\parbox[c]{2.25cm}{\centering MeanFlowTSE~\cite{shimizu25_meanflowtse}}}
      & w/o
        & 3.25 & 3.63 & 2.85 & 3.17
        & 1.51 & 0.60 & 0.03  & 0.57
        & \multirow[c]{2}{*}{\deltapct{-10.1\%}}
        & \multirow[c]{2}{*}{\deltapct{-31.7\%}}
        & \multirow[c]{2}{*}{\deltadb{-12.82}} \\
      &  & w/
        & 3.45 & 3.97 & 3.17 & 3.55
        & 2.21 & 0.82 & 12.85 & 0.73 \\
      \addlinespace[0.2em]

      & \multirow[c]{2}{*}{\parbox[c]{2.25cm}{\centering AlphaFlowTSE}}
      & w/o
        & 3.48 & 3.90 & 3.11 & 3.43
        & 2.16 & 0.82 & 12.76 & 0.76
        & \multirow[c]{2}{*}{\textbf{\deltapct{-2.5\%}}}
        & \multirow[c]{2}{*}{\textbf{\deltapct{-5.3\%}}}
        & \multirow[c]{2}{*}{\textbf{\deltadb{-0.40}}} \\
      &  & w/
        & \textbf{3.49} & \textbf{4.01} & \textbf{3.19} & \textbf{3.57}
        & \textbf{2.28} & \textbf{0.85} & \textbf{13.16} & 0.76 \\

    \bottomrule
  \end{tabular}
\end{table*}

\subsection{Evaluation Metrics}
\label{sec:metrics}

All metrics are computed on reconstructed waveforms at 16~kHz.
For Libri2Mix, where clean target references are available, we report both reference-based and non-intrusive metrics.
Specifically, we use wideband Perceptual Evaluation of Speech Quality (PESQ)~\cite{itu07_p8622} to assess perceptual quality, extended Short-Time Objective Intelligibility (ESTOI)~\cite{jensen16_estoi} to assess intelligibility, and scale-invariant signal-to-distortion ratio (SI-SDR)~\cite{leroux19_sisdr} to measure separation accuracy.
As a complementary non-intrusive measure, we compute the DNSMOS P.835 score (DNSMOS)~\cite{reddy21_dnsmos835} using the official implementation from the Microsoft DNS-Challenge repository~\cite{microsoft_dnschallenge}.
Speaker similarity is measured as cosine similarity between speaker embeddings extracted by a pretrained WeSpeaker encoder~\cite{wang22_wespeaker}; on Libri2Mix, the embeddings are computed from the extracted speech and the clean target reference.

For REAL-T PRIMARY~\cite{li25da_interspeech}, clean and time-aligned target references are not available, so reference-dependent metrics such as SI-SDR, PESQ, and ESTOI are not applicable.
Following the REAL-T evaluation protocol, we report transcript-based error rates: word error rate (WER) for English computed with Whisper-large-v2~\cite{radford22_whisper} and character error rate (CER) for Chinese computed with FireRedASR-AED-L~\cite{xu25_fireredasr}.
We additionally report speaker similarity between the extracted speech and the enrollment utterance using the pretrained WeSpeaker encoder~\cite{wang22_wespeaker}.

\section{Results}

\subsection{Libri2Mix: One-Step Benchmark Performance}
\label{sec:results_librimix}

Table~\ref{tab:librimix_overall} reports controlled one-step evaluation on Libri2Mix (\texttt{min}, 16~kHz), and Table~\ref{tab:librimix_mr_effect} compares the MR-enabled and predictor-free configurations.
All AlphaFlowTSE results use a single separator evaluation (NFE$=1$).

\noindent\textbf{Reference-based fidelity and intelligibility.}
Under the MR-enabled setting, AlphaFlowTSE records the highest PESQ, ESTOI, and SI-SDR among the compared one-step systems on both \texttt{clean} and \texttt{noisy} at separator NFE$=1$.
Relative to MeanFlowTSE, its margins in PESQ, ESTOI, and SI-SDR are 0.01, 0.01, and 0.37~dB on \texttt{clean}, and 0.07, 0.03, and 0.31~dB on \texttt{noisy}, with the same metric ordering across both splits.

\noindent\textbf{Perceptual quality and target-speaker similarity.}
AlphaFlowTSE records the highest DNSMOS OVRL among the compared one-step systems on both splits, while DDTSE and FlowTSE report higher OVRL on both \texttt{clean} and \texttt{noisy}.
For target-speaker similarity (SpkSim), AD-FlowTSE records the highest values on both splits; AlphaFlowTSE obtains 0.93 on \texttt{clean} and 0.76 on \texttt{noisy}, exceeding MeanFlowTSE's 0.73 on \texttt{noisy}.
These results separate the reference-based fidelity and intelligibility comparison from DNSMOS and SpkSim, for which the leading systems differ.

\begin{table}[t]
  \centering
  \caption{Separator evaluation count and model size comparison. NFE denotes the number of separator network function evaluations at test time. ``Params'' reports the separator (backbone) parameters, while ``Aux Params'' reports additional parameters required at inference (e.g., an MR predictor).}
  \label{tab:model_complexity}
  \footnotesize
  \setlength{\tabcolsep}{3.8pt}
  \renewcommand{\arraystretch}{1.08}
  \begin{tabular}{l c c c}
    \toprule
    Method & NFE & Params (M) & Aux Params (M) \\
    \midrule
    DiffSep+SV~\cite{zhang24_ddtse} & 60 & 66  & 6.63 \\
    DDTSE~\cite{zhang24_ddtse} & 10 & 71  & 6.63 \\
    SR-SSL~\cite{ku25_srssl}       & 5  & 431 & -- \\
    SoloSpeech~\cite{wang25_solospeech} & 50 & 589 & -- \\
    \midrule
    AD-FlowTSE~\cite{hsieh25_adflowtse}   & 1 or 5 & 342 & 15.57 (MR predictor) \\
    MeanFlowTSE~\cite{shimizu25_meanflowtse} & 1 & 343 & 15.57 (MR predictor) \\
    \textbf{AlphaFlowTSE} & \textbf{1} & 343 & Optional \\
    \bottomrule
  \end{tabular}
\end{table}
\begin{table*}[t]
  \centering
  \caption{DNSMOS OVRL on REAL-T PRIMARY~\cite{li25da_interspeech} (higher is better).
  Models are trained on Libri2Mix \texttt{noisy} and evaluated zero-shot on REAL-T PRIMARY.
  We report the predictor-free configuration (w/o MR predictor) and the MR-enabled configuration using a predictor imported from synthetic training (w/ MR predictor), since REAL-T provides no MR labels.}
  \label{tab:realt_dnsmos}
  \footnotesize
  \setlength{\tabcolsep}{2.7pt}
  \renewcommand{\arraystretch}{1.06}
  \begin{tabular}{l c ccc ccc}
    \toprule
    Dataset & $Samples(N)$ &
    \multicolumn{3}{c}{DNSMOS OVRL$\uparrow$ (w/o MR predictor)} &
    \multicolumn{3}{c}{DNSMOS OVRL$\uparrow$ (w/ MR predictor)} \\
    \cmidrule(lr){3-5}\cmidrule(lr){6-8}
    & & AD-FlowTSE & MeanFlowTSE & AlphaFlowTSE
      & AD-FlowTSE & MeanFlowTSE & AlphaFlowTSE \\
    \midrule
    \multicolumn{8}{l}{\textit{English subsets}} \\
    AMI        & 592  & 1.837 & \textbf{2.178} & 1.820 & 1.799 & 2.120 & \textbf{2.169} \\
    CHiME-6    & 545  & 1.460 & 1.174 & \textbf{1.843} & 1.597 & \textbf{1.896} & 1.858 \\
    DipCo      & 133  & 1.346 & 1.193 & \textbf{1.560} & 1.252 & 1.475 & \textbf{1.515} \\
    Avg.       & 1270 & 1.624 & 1.644 & \textbf{1.803} & 1.655 & 1.956 & \textbf{1.967} \\
    \midrule
    \multicolumn{8}{l}{\textit{Chinese subsets}} \\
    AISHELL-4  & 240  & \textbf{2.113} & 1.732 & 2.002 & 2.137 & 2.258 & \textbf{2.277} \\
    AliMeeting & 481  & 1.824 & 1.607 & \textbf{1.974} & 1.855 & 2.058 & \textbf{2.086} \\
    Avg.       & 721  & 1.921 & 1.648 & \textbf{1.983} & 1.949 & 2.125 & \textbf{2.150} \\
    \bottomrule
  \end{tabular}
\end{table*}

\subsection{MR Configurations and Evaluation Count}
\label{sec:results_mr_ablation}

Several recent one-step TSE baselines rely on an MR predictor to set a trajectory coordinate at inference.
To compare the two operating configurations, Table~\ref{tab:librimix_mr_effect} reports predictor-free and MR-enabled results together with their signed relative changes.

\noindent\textbf{Configuration comparison.}
AD-FlowTSE and MeanFlowTSE show large gaps between their predictor-free and MR-enabled configurations, particularly in SI-SDR for MeanFlowTSE.
The corresponding gaps for AlphaFlowTSE are smaller in SI-SDR, PESQ, and DNSMOS OVRL.
For AlphaFlowTSE, the MR-enabled configuration records higher PESQ, ESTOI, SI-SDR, and DNSMOS OVRL than the predictor-free configuration, while the differences between the two configurations are smaller than the corresponding differences for AD-FlowTSE and MeanFlowTSE.
Predictor-free AlphaFlowTSE records 18.50 and 12.76~dB SI-SDR on \texttt{clean} and \texttt{noisy}, compared with 19.17 and 13.16~dB for its MR-enabled configuration, characterizing both supported inference paths.

\noindent\textbf{Evaluation count and model size.}
Table~\ref{tab:model_complexity} summarizes separator NFE, backbone parameter count, and auxiliary parameters used at inference.
Iterative diffusion/cascaded systems require many separator evaluations, whereas the one-step family operates at NFE$=1$ for the separator.
Within one-step systems, MR-indexed baselines require an additional MR predictor (auxiliary parameters and an extra forward pass).
AlphaFlowTSE uses one separator evaluation in both operating configurations; the MR-enabled configuration adds one predictor forward pass, whose effect is reported in Table~\ref{tab:librimix_mr_effect}.
Thus, separator NFE counts calls to the extraction backbone, whereas the complete MR-enabled pipeline also includes the auxiliary predictor evaluation.

\subsection{REAL-T PRIMARY: Zero-Shot Transfer to Real Conversations}
\label{sec:results_realt}

We next evaluate zero-shot transfer on REAL-T PRIMARY, which contains real conversational mixtures without aligned clean targets.
All models are trained on Libri2Mix \texttt{noisy} and evaluated zero-shot on REAL-T PRIMARY.
REAL-T does not provide MR labels; therefore, MR prediction is not supervised on REAL-T.
We report both the predictor-free configuration and an MR-enabled configuration using a predictor imported from synthetic training to compare their cross-domain behavior.
This setting evaluates predictor-free transport without in-domain coordinate supervision and separately measures the behavior of an MR predictor transferred from synthetic data.

\begin{figure}[t]
  \centering
  \includegraphics[width=\columnwidth]{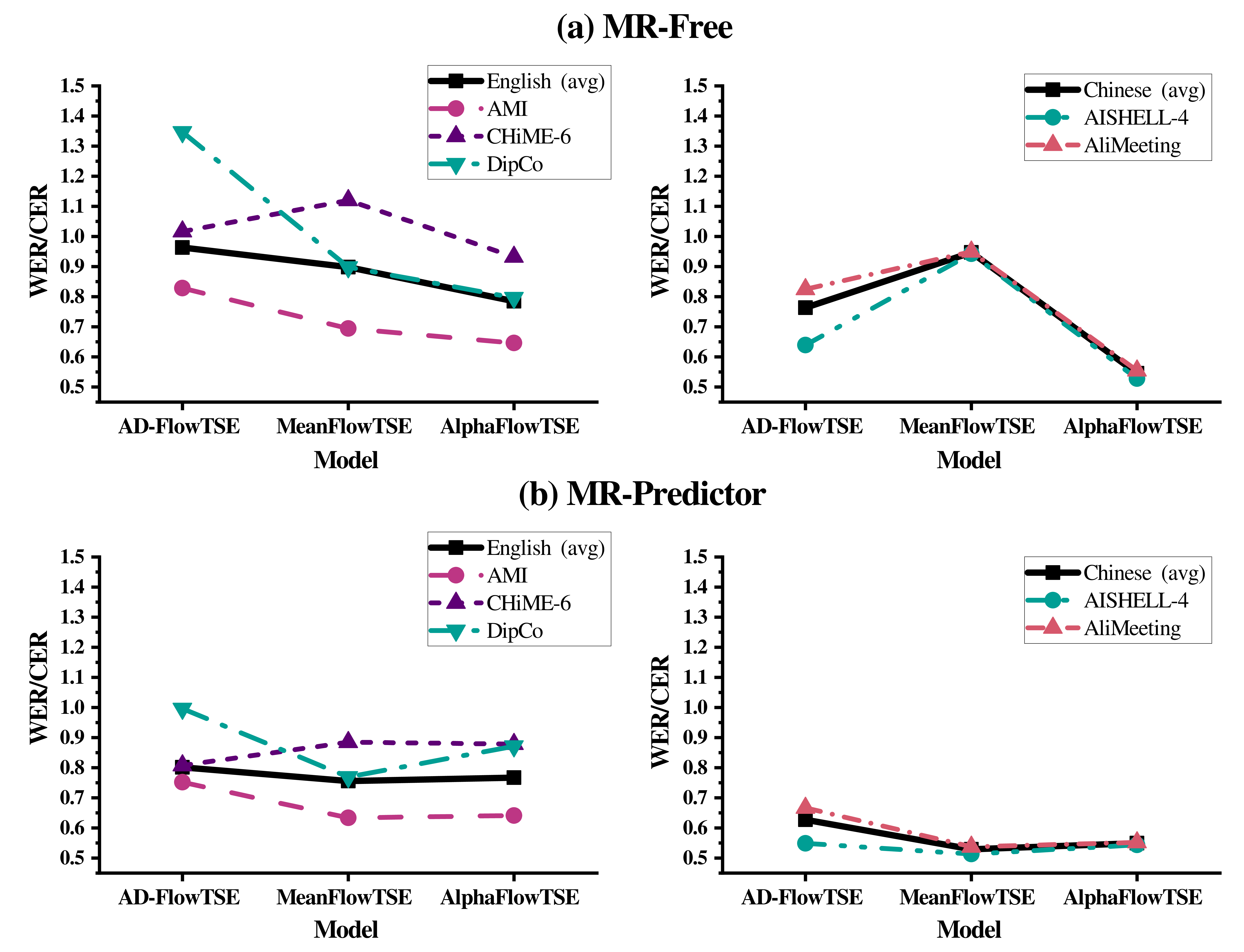}
  \caption{ASR error rates on REAL-T PRIMARY under two inference settings: (a) predictor-free (w/o MR predictor) and (b) MR-enabled (w/ MR predictor).
  Left panels report English WER (average and subsets: AMI, CHiME-6, DipCo), and right panels report Chinese CER (average and subsets: AISHELL-4, AliMeeting) for AD-FlowTSE, MeanFlowTSE, and AlphaFlowTSE.
  Lower is better.}
  \label{fig:realt_ter}
\end{figure}

\begin{figure}[t]
  \centering
  \includegraphics[width=\columnwidth]{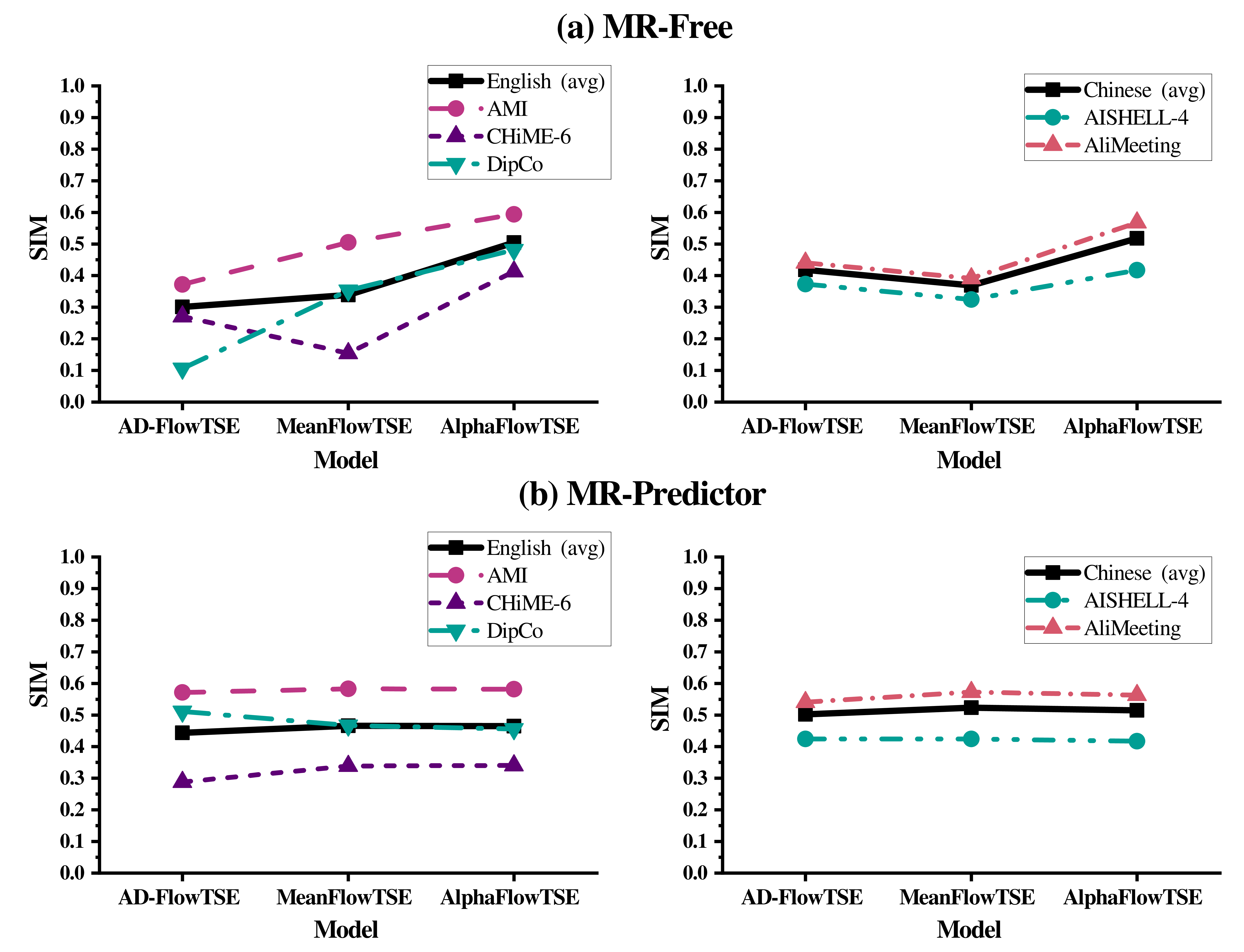}
  \caption{Speaker similarity (SIM / SpkSim) on REAL-T PRIMARY under two inference settings: (a) predictor-free (w/o MR predictor) and (b) MR-enabled (w/ MR predictor).
  Left panels report English SpkSim (average and subsets: AMI, CHiME-6, DipCo), and right panels report Chinese SpkSim (average and subsets: AISHELL-4, AliMeeting) for AD-FlowTSE, MeanFlowTSE, and AlphaFlowTSE.
  Higher is better.}
  \label{fig:realt_spksim}
\end{figure}

\noindent\textbf{Downstream ASR accuracy.}
Figure~\ref{fig:realt_ter} reports downstream ASR error rates on REAL-T PRIMARY.
Under predictor-free inference (Fig.~\ref{fig:realt_ter}(a)), AlphaFlowTSE yields the lowest WER/CER across all subsets and the lowest language-level averages.
Under the MR-enabled configuration (Fig.~\ref{fig:realt_ter}(b)), the lowest ASR error is achieved by different methods across subsets; AlphaFlowTSE records the lowest errors on every subset only in the predictor-free setting.
The contrast with Libri2Mix shows that configuration rankings differ between synthetic reference-based evaluation and cross-domain ASR.

\noindent\textbf{Target-speaker similarity.}
Under predictor-free inference (Fig.~\ref{fig:realt_spksim}(a)), AlphaFlowTSE achieves the highest SIM on both language averages and on all subsets.
Under the MR-enabled configuration (Fig.~\ref{fig:realt_spksim}(b)), the highest SIM is achieved by different methods across subsets.

\noindent\textbf{Reference-free perceptual quality.}
Table~\ref{tab:realt_dnsmos} reports DNSMOS OVRL on REAL-T PRIMARY.
AlphaFlowTSE achieves the highest DNSMOS OVRL on both language-level averages under both inference settings.
Under predictor-free inference, it ranks first on three of five subsets;
with the imported MR predictor, AlphaFlowTSE records higher language-level averages than under predictor-free inference and ranks first on four of five subsets.

\noindent\textbf{Summary.}
Across Libri2Mix and REAL-T PRIMARY, AlphaFlowTSE performs target-speaker extraction with separator NFE$=1$.
On REAL-T PRIMARY, its predictor-free configuration yields the lowest ASR errors and highest speaker similarity across all evaluated subsets; under the optional MR-enabled configuration, it records higher DNSMOS language-level averages, while the leading method for ASR and speaker similarity varies by subset.

\section{Conclusion}
We presented AlphaFlowTSE, a one-step generative TSE framework that learns mean-velocity mixture-to-target transport and is trained with a JVP-free AlphaFlow objective combining trajectory matching and interval-consistent teacher--student supervision. On Libri2Mix, the MR-enabled configuration achieves the highest PESQ, ESTOI, and SI-SDR among the compared one-step systems, while the gap between its predictor-free and MR-enabled results is smaller than in the corresponding baseline comparisons. On REAL-T PRIMARY, predictor-free inference yields the lowest ASR errors and highest speaker similarity across the evaluated subsets. Together, the two configurations provide complementary operating points across synthetic and real conversational evaluations.

\section{Acknowledgements}
This research was supported in part by the National Natural Science Foundation of China under Grant No. 62271432 and in part by the Program for Guangdong Introducing Innovative and Entrepreneurial Teams under Grant No. 2023ZT10X044.

\bibliographystyle{IEEEtran}
\bibliography{mybib}

\end{document}